\documentclass[10pt,letter]{article}

\usepackage{graphicx} 
\usepackage{url}

\usepackage{hyperref}
\usepackage{natbib}
\usepackage{times}

\def\ITMProbe{\textsf{{\footnotesize\itshape ITM Probe}}}

\begin{document}

\begin{titlepage}

\begin{center}
{\Large\bf ITM Probe: analyzing information flow in protein networks}
\end{center}
\vspace{.35cm}

\begin{center}
{\large Aleksandar Stojmirovi\'c\, and Yi-Kuo Yu\footnote{to whom correspondence should be addressed}}
\vspace{0.25cm}
\small

\par \vskip .2in \noindent
National Center for Biotechnology Information\\
National Library of Medicine\\
National Institutes of Health\\
Bethesda, MD 20894\\
United States
\end{center}

\normalsize
\vspace{0.25cm}

\begin{abstract}

\subsubsection*{Summary:}
Founded upon diffusion with damping, \ITMProbe\ is an application for modelling information flow in protein interaction networks without prior restriction to the sub-network of interest. Given a context consisting of desired origins and destinations of information, \ITMProbe\ returns the set of most relevant proteins with weights and a graphical representation of the corresponding sub-network. With a click, the user may send the resulting protein list for enrichment analysis to facilitate hypothesis formation or confirmation.

\subsubsection*{Availability:}
\ITMProbe\ web service and documentation can be found at \href{www.ncbi.nlm.nih.gov/CBBresearch/qmbp/mn/itm\_probe}{www.ncbi.nlm.nih.gov/CBBresearch/qmbp/mn/itm\_probe}

\subsubsection*{Contact:} \href{yyu@ncbi.nlm.nih.gov}{yyu@ncbi.nlm.nih.gov}
\end{abstract}
\end{titlepage}

\section{Introduction}

Protein interaction networks are presently under intensive research \citep{BKG08}. Recently, a number of authors have applied the concept of random walk (with truncation) to extract biologically relevant information from protein interaction networks \citep{NJAC05,TWACS06,SBKEI08}. These approaches, however, do not model information loss/leakage that naturally occurs in all networks. For example, in cellular networks, proteases constantly degrade proteins, diminishing the strength of information propagation. We have recently developed a mathematical framework to model information flow in interaction networks with a novel ingredient, damping/aging of information \citep{SY07}. Implementing the theory, we have constructed  a web application \ITMProbe, which also contains a new model of information propagation: information channel.

\ITMProbe\ models information flow in a protein interaction network through discrete random walks. Unlike classical random walks, our model allows the walker a certain probability to \emph{dissipate} or \emph{damp} (that is, to leave the network) at each step. Each walk, simulating a possible information path, terminates either by dissipation or by reaching a boundary node.

We distinguish two types of boundary nodes: \emph{sources} (emitting information) and \emph{sinks} (absorbing information). \ITMProbe\ offers three models: absorbing, emitting and channel. For any network node, the corresponding weight returned by the emitting model is the expected number of visits to that node by a random walk originating at given source(s). The absorbing model, on the other hand, returns the likelihood of a random walk starting at that node to terminate at sink(s). The channel model combines the emitting and absorbing models: it contains both sources and sinks as boundary and reports the expected numbers of visits to any network node from random walks originating at sources and terminating at sinks.

Each selection of boundary nodes and dissipation rates provides the biological \emph{context} for the information transmission modelled. Small dissipation allows random walks to explore the nodes farther away from their origin while large dissipation evaporates quickly most walks. For the channel model, dissipation controls how much a random walk can deviate from the shortest path from sources to sinks. We call the set of most significant nodes, in terms of the weights returned, an \emph{Information Transduction Module} (ITM).

\section{Usage}\label{sec:usage}

Both the absorbing and emitting models navigate neighborhoods of selected nodes and illuminate the protein complexes associated with them. However, the absorbing model can reveal relatively distant `leaf' nodes linked to a sink by a nearly unique path, while the emitting model favors highly connected clusters. The channel model is suited for discovery of potential pathways linking proteins of interest or biological functions associated with them. Using multiple sources may reveal the potential points of crosstalk between information channels, while a solution of multiple sinks chosen according to a set of competing hypotheses may suggest the most biologically plausible pathways among many possible ones.

Every model of \ITMProbe\ requires an interaction graph, the boundary nodes (sources and/or sinks) and the damping factors as input. The damping factors may be specified directly or by setting the desired average path-length (emitting/channel model) or the average likelihood of absorption at sinks (absorbing model). 

Although our mathematical framework can be applied to any directed graph, our web service presently supports only the yeast (\textit{Saccharomyces cerevisiae}) physical interaction networks derived from the BioGRID \citep{SBRBT06} database. We offer three yeast networks: Full, Reduced and Directed. The Full network consists of all interactions from the BioGRID as an undirected graph, while the Reduced consists only of those interactions that are from low-throughput experiments (that is, from publications reporting less than $300$ interactions) or are reported by at least two independent publications. The Directed network is derived from Reduced by turning all interactions labelled as `Biochemical activity' into directed links (bait $\to$ prey).

To assist \textit{in silico} investigations on the impact of knocking out certain genes, \ITMProbe\ allows users to specify nodes to exclude from the network. Furthermore, it is known \citep{SPAD02} that proteins with a large number of non-specific interaction partners might overtake the true signaling proteins in the information flow modeling. Therefore, \ITMProbe\ by default excludes from the yeast networks the proteins that may provide undesirable shortcuts, such as cytoskeleton proteins, histones and chaperones. The user may choose to lengthen or shorten this list.

\subsection*{Output and analysis}
\begin{figure}[t]
\begin{center}
\scalebox{0.7}{\includegraphics{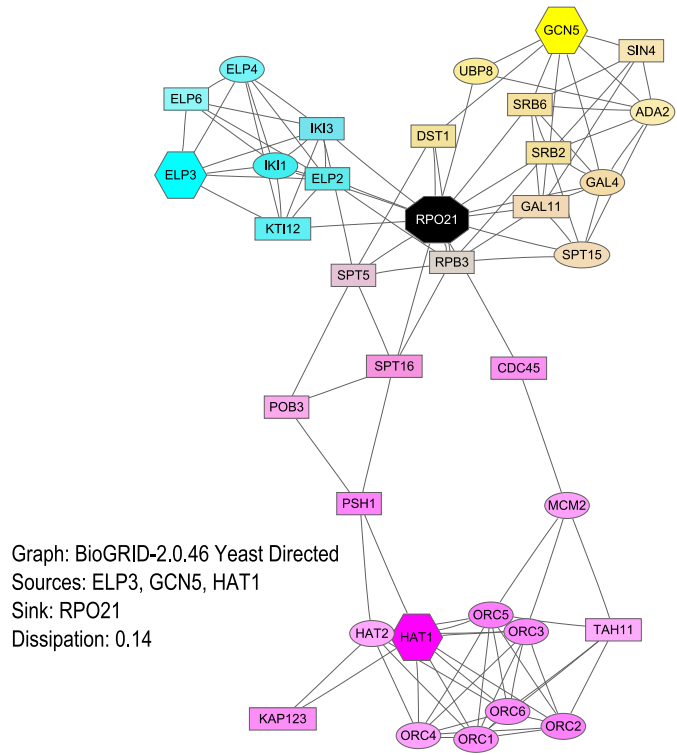}}
\caption{An example ITM from running the \ITMProbe\ channel model.} \label{fig:screenshot}
\end{center}
\end{figure}

\ITMProbe\ outputs a list of the top ranking nodes together with an image of the sub-network consisting of these nodes (Fig.~\ref{fig:screenshot}). Images are produced using the Graphviz suite \citep{GN00}. Each protein listed is linked to its full description in several external databases. The number of nodes to be listed can be specified directly by the user or determined automatically from the model results through a criterion such as participation ratio \citep{SY07} or the cutoff value. The  resulting weights for all nodes can be downloaded in the CSV format for further analysis.

Each ITM image can be rendered and saved in multiple formats (SVG, PNG, JPEG, EPS and PDF). For each rendering, the users can choose which aspects of results to display, the color map and the scale for presentation (linear or logarithmic). When multiple boundary points are specified, it is possible to obtain an overview of all of their contributions simultaneously by selecting the color mixture scheme (Fig.~\ref{fig:screenshot}). In this case, each source (channel/emitting model) or sink (absorbing model) is assigned a basic CMY (cyan, magenta or yellow) color and the coloring of each displayed node is a result of mixing the colors corresponding to its source- or sink- specific values for each of the boundary points. 

While it is possible to specify any proteins in the network as sources and sinks, not every context produces biologically meaningful results. To facilitate biological interpretation of the users' results, we have locally implemented a Gene Ontology (GO) \citep{ABB00} enrichment tool based on \textsf{GO::TermFinder} of \citet{BWGJBCS04}. It compares a given input list of proteins to the lists annotated with GO terms and finds those GO terms that statistically best explain the input list. Every \ITMProbe\ results page contains a query form allowing the user to specify the number of the top ranking proteins to consider for GO term enrichment analysis.

\subsection*{Example}

Histone acetyltransferases remodel chromatin by acetylating histone octamers and hence may play an important role in transcription activation \citep{SB00}. To explore the interface between them and the RNA Polymerase II core in yeast, we choose three histone acetyltransferases (Hat1p, Gcn5p, Elp3p) as sources and a catalytic subunit Rpo21p of RNA Polymerase II as a sink for the channel model (Fig.~\ref{fig:screenshot}). From the color mixing image it appears that Elp3p and Gcn5p interact with Rpo21p through a wide channel of proteins, while Hat1p seems to be remote from Rpo21p. This prompts the hypothesis that Hat1p is not directly involved in transcription activation. Enrichment analysis, using the 16 nodes (shown in magenta color in Fig.~\ref{fig:screenshot}) mostly visited from Hat1p, shows that Hat1p and these nodes participate mainly in DNA replication and only indirectly in transcription regulation, thus reinforcing the hypothesis. Similar analysis on the nodes associated with Elp3p indicates the interaction is almost exclusively through the elongator complex. The nodes associated with Gcn5p are less specific, indicating a more generic interface, but are all involved mRNA transcription.

\section{Outlook}\label{sec:conclusion}
We plan to include interaction networks from additional organisms, once their coverage/quality becomes comparable to those from yeast. In principle, the analysis from \ITMProbe\ can be integrated with existing partial knowledge to form a broad picture of possible communication paths in cellular processes. The concept of context-specific analysis may find applications beyond biological networks.

\section*{Acknowledgments}

This work was supported by the Intramural Research Program of the National Library of Medicine at National Institutes of Health. \ITMProbe\ implementation relies on a variety of open source projects, which we acknowledge on our website.

\end{document}